\documentclass[sigconf]{acmart}
\AtBeginDocument{%
  }

\setcopyright{acmlicensed}
\copyrightyear{2018}
\acmYear{2018}
\acmDOI{XXXXXXX.XXXXXXX}
\acmConference[Conference acronym 'XX]{Make sure to enter the correct
  conference title from your rights confirmation email}{June 03--05,
  2018}{Woodstock, NY}
\acmISBN{978-1-4503-XXXX-X/2018/06}

\begin{document}

\title{HHFT: Hierarchical Heterogeneous Feature Transformer for Recommendation Systems}

\author{Liren Yu}
\affiliation{%
  \institution{Taobao $\&$ Tmall Group of Alibaba}
  \city{Hangzhou}
  \country{China}
}
\email{yuliren.ylr@taobao.com}

\author{Wenming Zhang}
\affiliation{%
 \institution{Taobao $\&$ Tmall Group of Alibaba}
  \city{Hangzhou}
  \country{China}}
\email{zhangwenming.zwm@taobao.com}

\author{Silu Zhou}
\affiliation{%
 \institution{Taobao $\&$ Tmall Group of Alibaba}
  \city{Hangzhou}
  \country{China}
}
\email{silu.zsl@taobao.com}

\author{Tao Zhang}
\affiliation{%
 \institution{Taobao $\&$ Tmall Group of Alibaba}
  \city{Beijing}
  \country{China}
}
\email{quen.zt@alibaba-inc.com}

\author{Zhixuan Zhang}
\affiliation{%
 \institution{Taobao $\&$ Tmall Group of Alibaba}
  \city{Hangzhou}
  \country{China}}
  \email{zhibing.zzx@taobao.com}

\author{Dan Ou}
\affiliation{%
 \institution{Taobao $\&$ Tmall Group of Alibaba}
  \city{Hangzhou}
  \country{China}}
  \email{oudan.od@taobao.com}

\renewcommand{\shortauthors}{Trovato et al.}

\begin{abstract}
   We propose HHFT (Hierarchical Heterogeneous Feature Transformer), a Transformer-based architecture tailored for industrial CTR prediction. HHFT addresses the limitations of DNN through three key designs: (1) Semantic Feature Partitioning: Grouping heterogeneous features (e.g. user profile, item information, behaviour sequennce) into semantically coherent blocks to preserve domain-specific information; (2) Heterogeneous Transformer Encoder: Adopting block-specific QKV projections and FFNs to avoid semantic confusion between distinct feature types; (3) Hiformer Layer: Capturing high-order interactions across features.
  Our findings reveal that Transformers significantly outperform DNN baselines, achieving a $+0.4\%$ improvement in CTR AUC at scale. We have successfully deployed the model on Taobao's production platform, observing a significant uplift in key business metrics, including a $+0.6\%$ increase in Gross Merchandise Value (GMV).
\end{abstract}




\maketitle

\section{Introduction}

Recommendation systems serve as the cornerstone of modern digital ecosystems, particularly in e-commerce platforms, where they directly influence user engagement and business revenue through personalized content delivery. Over the past decade, recommendation technologies have evolved from traditional collaborative filtering to deep learning-based approaches, with DNNs becoming the de facto standard for CTR prediction tasks \cite{covington2016deep,cheng2016wide,guo2017deepfm,zhou2018deep,wang2021dcn}. However, despite their universal approximation capabilities \cite{mhaskar1996neural}, DNN-based methods still struggle to explicitly model high-order feature interactions in sparse, high-dimensional user-item-context data. Their feedforward architectures fail to preserve semantically meaningful cross-feature relationships \cite{beutel2018latent,wang2021dcn}. 

This limitation is especially evident in e-commerce search scenarios, where ranking performance depends on the nuanced interplay of user intent (e.g., search queries), item attributes (e.g., price, category), and contextual factors (e.g., time, device). Classic DNN-based models such as Wide$\&$Deep\cite{cheng2016wide}, DeepFM\cite{guo2017deepfm} and DCNv2\cite{wang2021dcn} attempt to address this issue through manual feature engineering or implicit interaction modeling, but they still struggle with scalable high-order interaction learning and semantic preservation for heterogeneous features.

The Transformer architecture, originally proposed in natural language processing (NLP) \cite{vaswani2017attention}, has introduced a paradigm shift in sequence modeling through its self-attention mechanism, which dynamically captures pairwise dependencies between input elements. This capability enables adaptive and interpretable feature interaction learning \cite{vaswani2017attention}, making it particularly suitable for recommendation tasks.Furthermore, its predictable scaling behavior \cite{kaplan2020scaling} offers a clear path for performance gains in data-rich industrial settings. While recent studies have explored Transformers for recommendations, critical gaps remain. For example, SASRec \cite{kang2018self} and LONGER \cite{chai2025longer} leverage the Transformer structure to capture long-term dependencies in user behavior sequences. However, these approaches primarily focus on sequential data and overlook broader heterogeneous feature interactions. Hiformer \cite{gui2023hiformer} explicitly model heterogeneous feature interactions but lack a rigorous empirical analysis of scaling laws. Wukong \cite{zhang2024wukong} and Rankmixer \cite{zhu2025rankmixer} explore the potential of scaling laws for explicit high-order feature interaction in recommendations, but rely on non-Transformer elements (e.g., Factorization Machine blocks in Wukong or MLP-mixer \cite{tolstikhin2021mlp} in Rankmixer), limiting their ability to fully exploit Transformer's semantic-aware capabilities.

To bridge these gaps, this paper proposes Hierarchical Heterogeneous Feature Transformer
(HHFT) tailored for CTR prediction in ranking systems. HHFT addresses two core limitations of DNN-based rankers: (1) Explicit High-Order Interaction Modeling: By integrating semantic feature partitioning and heterogeneous attention mechanisms, HHFT directly models high-order interactions through learnable affinity matrices, overcoming the implicit and compressed feature fusion of DNNs. (2) Scalable Representation Learning: We validate that HHFT adheres to scaling laws, enabling predictable performance gains with increased model capacity.

Our key contributions are threefold:
\begin{enumerate}
\item  Architectural Innovation: We propose a hierarchical heterogeneous feature processing framework that partitions features by semantics and maintains domain-specific parameters, preserving the uniqueness of heterogeneous features while enabling cross-domain interactions.
\item Empirical Validation of Scaling Laws: We establish and validate predictable scaling relationships between model size and CTR prediction performance, providing a quantitative guide for model scaling in industrial settings.
\item Industrial Deployment and Business Impact: We successfully deploy HHFT on Taobao's production search platform, achieving statistically significant improvements in both model performance (AUC) and business metrics (GMV, CTR), demonstrating its practical application value.
\end{enumerate}

\section{Related Work}

\subsection{DNN-Based CTR Models}
DNN-based methods dominate CTR prediction by modeling nonlinear interactions. Wide$\&$Deep \cite{cheng2016wide} combines a wide linear layer (memorization) with a deep network (generalization). DeepFM \cite{guo2017deepfm} integrates Factorization Machines (FMs) into DNNs to model low-order interactions explicitly. DCNv2 \cite{wang2021dcn} uses cross layers to learn high-order interactions via feature cross operations. These models, however, rely on implicit interaction learning and struggle with heterogeneous features.
\subsection{Transformer-Based Recommender Systems}
Transformers have gained traction for their explicit attention mechanisms. SASRec \cite{kang2018self} and LONGER \cite{chai2025longer} apply Transformer encoders to user behavior sequences, capturing long-term dependencies. AutoInt \cite{song2019autoint} replaces DNN layers with self-attention to model feature interactions, but uses shared parameters for all features. Hiformer \cite{gui2023hiformer} proposes heterogeneous attention for feature interactions, but lacks scaling law analysis. 
\subsection{Scaling Laws in Recommendation}
Scaling laws, where performance improves predictably with model size, are critical for industrial platforms. Recent works like Wukong \cite{zhang2024wukong} RankMixer \cite{zhu2025rankmixer} explore scaling laws for recommendation, but Wukong relies on FM blocks and RankMixer designs token-mixing from mlpmixer \cite{tolstikhin2021mlp}  (not Transformers).

Scaling laws—where model performance improves in a predictable manner with increased model size—are of critical importance for industrial recommendation platforms. Recent works focusing on scaling laws for recommendation tasks include Wukong \cite{zhang2024wukong} and RankMixer \cite{zhu2025rankmixer}. However, Wukong relies on Factorization Machine (FM) blocks for feature interaction modeling, and RankMixer designs its token-mixing mechanism based on MLP-mixer \cite{tolstikhin2021mlp} instead of adopting Transformer-based attention and further ignores heterogeneous feature partitioning. These limitations mean that neither one of the solutions fully utilizes the strengths of the Transformer to capture complex, semantically aware feature interactions— a key gap addressed by our proposed HHFT.

This paper is structured as follows: Section 2 details the architectural of Transformer-based CTR ranking model; Section 3 presents scaling experiments  and online results; Section 5 explores implications and future work.

\section{Methodology}
\label{sec:model}

\subsection{Overview}
\begin{figure}[!ht]
    \centering
    \includegraphics[width=0.45\textwidth]{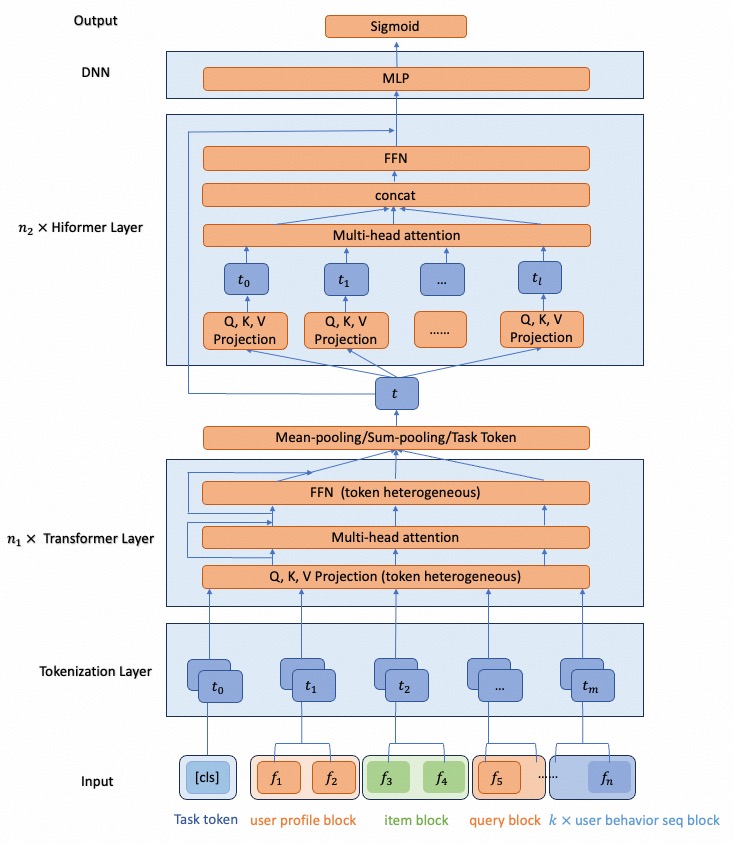}
    \caption{HHFT Architecture.}
    \label{fig:hhft_arch}
\end{figure}

Our proposed \textbf{Hierarchical Heterogeneous Feature Transformer (HHFT)}  replaces traditional DNN backbones with a Transformer-based structure optimized for recommendation scenarios. As shown in Figure~\ref{fig:hhft_arch}, HHFT processes input features through five stages:
\begin{enumerate}
    \item \textbf{Feature Partitioning}: Group raw features into semantically coherent blocks to preserve domain-specific information.
    \item \textbf{Heterogeneous Feature Tokenization}: Convert each feature block into  unified dimension tokens through embedding and projection, enabling inter-block computation.
    \item \textbf{Heterogeneous Transformer Encoder}: Model interactions between feature tokens using Transformer with domain-specific parameters.
    \item \textbf{Hiformer Layer\cite{gui2023hiformer}}: Enhance high-order interaction across all features through a sophisticated attention mechanism.
    \item \textbf{Prediction Head}: Generate final CTR/CVR predictions using a MLP.
\end{enumerate}

\subsection{Feature Tokenization}

The features of E-commerce recommendation/search system exhibit high heterogeneity, including categorical features (e.g., user ID, item category), continuous features (e.g., item price, user purchase frequency) and sequential features (e.g., user recent behavior). Directly concatenating these features leads to semantic information loss. Thus, we first partition the input features into $K$ disjoint blocks based on semantic categories:
\begin{equation}
\mathcal{B} = \{ \underbrace{\mathbf{U}}_{\text{user features}}, \underbrace{\mathbf{I}}_{\text{item features}}, \underbrace{\mathbf{Q}}_{\text{query features}}, \underbrace{\mathbf{S}}_{\text{behavior sequence}}, ... \}
\end{equation}

Each block $\mathbf{B}_k \in \mathbb{R}^{d_k}$ has a different dimension. Through embedding layers:
\begin{equation}
\mathbf{E}_k = \mathrm{Embed}_k(\mathbf{B}_k) \in \mathbb{R}^{e_k}
\end{equation}
where $e_k$ is the size of the block-specific embedding.

To enable cross-block attention, we project all embedding vectors to a unified dimension  $d$ using block-specific linear layers:
\begin{equation}
\mathbf{H}_k^{(0)} = \mathbf{W}_k^{\text{proj}} \mathbf{E}_k + \mathbf{b}_k^{\text{proj}}, \quad \mathbf{W}_k^{\text{proj}} \in \mathbb{R}^{d }
\end{equation}
yields an aligned token matrix $\mathbf{H}^{(0)} = \left[\mathbf{H}_U, \mathbf{H}_I, \mathbf{H}_Q, \mathbf{H}_S,...\right]^T \in \mathbb{R}^{K \times d}$.

\subsection{Heterogeneous Transformer Encoder}
\label{subsec:hetero_transformer}

Traditional Transformer encoders adopt shared parameters for all input tokens, which fails to preserve the uniqueness of heterogeneous features (e.g., categorical user IDs, continuous item prices, and sequential behavior features) in e-commerce recommendation/search scenarios. This deficiency often leads to semantic information confusion and suboptimal interaction modeling. To address this issue, the Heterogeneous Transformer Encoder proposed in this paper maintains independent parameter sets for Query (Q), Key (K), Value (V) projection and Feed-Forward Network (FFN) for each feature block (user/item/query/behavior sequence...), while reusing the core attention computation logic to ensure both feature specificity and computational efficiency. The detailed implementation is as follows:

\begin{enumerate}
\item \textbf{Block-Specific QKV Projection} For $k$-th feature block at the $l$-th encoder layer, the token representation $(H_k^{(l)} \in \mathbb{R}^d)$ is mapped to block-specific Q, K, V vectors using independent projection matrices. The calculation formula is: $Q_k = W_{Q,k} \cdot H_k^{(l)} , \quad K_k = W_{K,k} \cdot H_k^{(l)} , \quad V_k = W_{V,k} \cdot H_k^{(l)}$, 
where $W_{Q,k}, W_{K,k}, W_{V,k} \in \mathbb{R}^{d \times d}$  are block-specific weight. This design ensures that the attention weight calculation adapts to the semantic characteristics of each feature domain.

\item \textbf{Multi-Head Self-Attention } To capture high-order feature interactions, the Q, K, V vectors of each block are processed by standard multi-head self-attention.

\item \textbf{Block-Specific FFN} After attention computation, each block's token representation is input to a block-specific FFN to enhance nonlinear feature transformation capability. The FFN adopts a two-layer fully connected structure with ReLU activation, and the formula is:
 $FFN_k(x) = \text{ReLU}(x \cdot W_{1,k} + b_{1,k}) \cdot W_{2,k} + b_{2,k}$ where $W_{1,k} \in \mathbb{R}^{d \times d},b_{1,k} \in \mathbb{R}^{4d},W_{2,k} \in \mathbb{R}^{d \times d},b_{2,k} \in \mathbb{R}^d$ are block-specific parameters. The hidden layer dimension is set to $d$ to balance expressive power and computational cost.
\end{enumerate}

\subsection{Hiformer Layer}

Building on the heterogeneous Transformer encoder, the Hiformer layer\cite{gui2023hiformer} introduces composite projections to model interactions beyond pairwise dependencies.We use a global composite projection:
\begin{equation}
   [\hat{K}^h_1, \dots, \hat{K}^h_k] = \text{concat}([H^h_1, \dots, H^h_k]) \hat{W}^h 
\end{equation}
where  $\hat{W}^h \in \mathbb{R}^{k d \times k d_h}$. The same projection strategy is applied to $Q$ and $V$, ensuring symmetry in modeling interactions.
This design enables Hiformer to learn more comprehensive hierarchical interactions  across features.

\subsection{Prediction Head}
After $n_1$ transformer layers and $n_2$ hiformer layers, we concatenate all tokens and generate the final CTR/CVR prediction via a MLP.

\section{Experiment}
To comprehensively validate the effectiveness, component contributions, scaling properties, and industrial applicability of HHFT, we conduct experiments on Taobao’s real-world e-commerce data.

\subsection{Experiment Settings}
\subsubsection{Datasets and Environment.} To evaluate HHFT, we conducted both offline and online experiments on Taobao's e-commerce dataset, comprising billions of user-item interactions.
\subsubsection{Evaluation Metrics.} To evaluate model performance, we use AUC (Area Under the Curve) as the primary performance metrics. An AUC increase of 0.001 can be regarded as a confidently significant improvement.  We also report model size (Params, in millions) and computational complexity (TFLOPs) to evaluate efficiency-scalability tradeoffs. For online experiments, we use GMV as the business-centric metric to reflect practical impact.
\subsubsection{Baselines.} We compare against the following widely recognized SOTA baselines: 
\textbf{DLRM-MLP}:which is the vanilla MLP for
feature crossing as the experiment baseline; \textbf{DCNv2}\cite{wang2021dcn}
the sota of feature cross model.
\textbf{AutoInt}\cite{song2019autoint}, \textbf{Hiformer}\cite{gui2023hiformer} investigate transform-like architecture for ranking model.
\textbf{Wukong}\cite{zhang2024wukong} and \textbf{Rankmixer}\cite{zhu2025rankmixer} investigate the scaling law  for recommendation system.

\subsection{Comparison with SOTA methods}

The main results are summarized in Table \ref{tab:cmp_sota}. HHFT outperforms all SOTA baselines in both predictive accuracy (AUC) and scalability, aligning with our design goals. We can also observe that Transformer-based models (AutoInt, Hiformer, HHFT) outperform DNN/FM-based models (DCNv2, Wukong), confirming the value of explicit attention for feature interactions. Furthermore, HHFT’s AUC gain (vs. DLRM-MLP) exceeds Hiformer and Wukong, attributed to its hierarchical heterogeneous design and scaling optimization.

\begin{table}[h]
\centering
\caption{Performance comparison of recommendation models (best values in bold)
}
\scalebox{1}{
\begin{tabular}{c|c|c|c}
\hline
Model &  AUC & Params(M) & TFLOPs \\
\hline
 DLRM-MLP (base) & -  &15 &0.42 \\ 
 DCNv2 &  +0.001 &24 &0.65\\ 
AutoInt  & +0.005 &150 & 1.19\\
HiFormer  & +0.005 & 170& 1.98\\ 
Wukong  & +0.002 & 32& 0.94\\ 
Rankmixer  & +0.003 & 140& 1.93\\ 
HHFT & \textbf{+0.008}& 300& 1.22\\ 
\hline
\end{tabular}}
\label{tab:cmp_sota}
\end{table}

\subsection{Ablation Study}

To quantify the incremental contribution of each key component in HHFT, we conduct ablation experiments under a cold-start training scenario (with 7 consecutive days of cold-start training, where all dense layer weights initialized randomly).  Table \ref{tab:cmp_arch} presents the AUC gains relative to the DLRM-MLP baseline, with key insights as follows:

\begin{enumerate}
    \item Transformer Architecture: Replacing DNN-MLP with a basic Transformer encoder already yields significant gains, confirming that self-attention outperforms MLP in modeling feature interactions.
    \item Heterogeneous Parameterization Introducing token-specific QKV projections and FFNs further improves performance, validating that avoiding semantic confusion between heterogeneous features is critical for interaction modeling.
    \item Hiformer Layer: Adding Hiformer layers enhances high-order interactions, demonstrating the value of moving beyond pairwise dependencies.
    \item Weight Initialization Optimization: Tuning parameter initialization and hyperparameters (e.g., layer normalization placement, learning rate scheduling) delivers the large gain, highlighting the importance of industrial-friendly training optimizations for Transformer-based models.
    \item Scaling Up: Scaling up model size improves performance, consistent with scaling laws.
\end{enumerate}

The combined effect of these components results in a total +0.0117 AUC gain, confirming that each design choice in HHFT contributes meaningfully to its performance advantage.

\begin{table}[h]
\centering
\caption{Ablation on components of HHFT.
}
\scalebox{1}{
\begin{tabular}{c|c}
\hline
Setting &  auc gain vs MLP  \\
\hline
 MLP $\to$ Transformer & +0.0035  \\ 
 Heterogeneous Transformer & +0.0018 \\ 
Hiformer Layer & +0.0011 \\
Weight Initialization  & +0.0040\\ 
Scaling up  & +0.0034 \\ 
\hline
\end{tabular}}
\label{tab:cmp_arch}
\end{table}

\subsection{Scaling Laws Validation}

A key advantage of Transformer-based models is compliance with scaling laws—performance improves predictably with model size. We define the "base model" of HHFT with the following dense parameters (excluding embeddings, which are fixed): 
\begin{itemize}
    \item Heterogeneous Transformer: layer ($n_1=1$), token dimension ($d_\text{trfm}=1648$), FFN dimension ($d_\text{ffn}=1648$)
    \item Hiformer Layer: layer ($n_2=1$), token dimension ($d_\text{hifm}=256$), number of tokens ($n_h=8$)
\end{itemize}

\begin{figure}[!ht]
    \centering
    \includegraphics[width=0.36\textwidth]{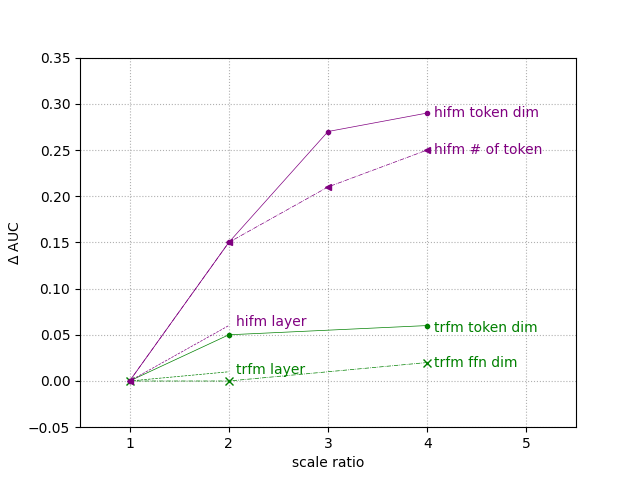}
    \caption{AUC gain vs Dense Parameters Scale Ratio}
    \label{fig:scaling_exp}
\end{figure}

Then, we scale each parameter independently by a factor (keeping other parameters fixed), and measure AUC gain in Figure \ref{fig:scaling_exp} Two critical conclusions emerge for industrial model scaling:
\begin{enumerate}
    \item Width > Depth: Scaling the width of the model (e.g., Transformer/Hiformer token dimension) yields  higher AUC gains than scaling depth (number of Transformer/Hiformer layers). This provides a practical guide for resource-efficient model expansion.
     \item High-Order > Low-Order Scaling: Scaling parameters related to high-order interactions (e.g., Hiformer token count and token dimension) is more effective than scaling low-order components (e.g., Transformer layers). This confirms that HHFT’s focus on hierarchical high-order interactions is not only architecturally innovative but also scaling-efficient.
\end{enumerate}

\subsection{Online Performance}
To validate HHFT’s industrial practicality, we conducted a 30-day A/B test on Taobao’s search platform, allocating $1\%$ of total traffic to the HHFT group (with DNN as the control baseline). Results show HHFT outperforms the DNN baseline by $+0.4\%$ in CTR AUC—an indicator reflecting accurate user click intent distinction—and drives a $+0.6\%$ increase in GMV. For Taobao’s massive user base and transaction volume, these gains translate to substantial incremental revenue, fully confirming the model’s real-world efficacy .

\section{Conclusion and Future Work}
This paper proposes HHFT, a hierarchical heterogeneous Transformer for e-commerce CTR prediction. By combining semantic feature partitioning, domain-specific parameters Transformer, and Hiformer layers, HHFT outperforms SOTA models and delivers significant online business gains. We also validate HHFT’s compliance with scaling laws, providing a reliable path for industrial ranking model scaling.

Future work will focus on extending HHFT to joint ranking for search, recommendation, and advertising—sharing cross-domain features to further improve business value.



\bibliographystyle{ACM-Reference-Format}
\bibliography{sample-base}


\end{document}